\documentclass[prl,reprint, twocolumn,sort&compress,superscriptaddress]{revtex4-1}

\usepackage{lineno,hyperref}
\usepackage{graphicx}
\usepackage{amsmath}
\usepackage{epsfig}
\usepackage{subfigure}
\usepackage{tabularx,ragged2e}
\usepackage[detect-all]{siunitx}
\usepackage{rotating}

\begin{document}

\title{Effect of fractional blood flow on plasma skimming in the microvasculature}

\author{Jiho Yang}
\affiliation{Advanced Institutes of Convergence Technology, Seoul National University, Suwon 443-270, Republic of Korea}
\affiliation{Department of Computer Science, Technische Universit{\"a}t M{\"u}nchen, Boltzmannstra{\ss}e 3, Garching, Germany}
\author{Sung Sic Yoo}
\affiliation{Advanced Institutes of Convergence Technology, Seoul National University, Suwon 443-270, Republic of Korea}
\author{Tae-Rin Lee}
\thanks{Corresponding author, taerinlee@snu.ac.kr}
\affiliation{Advanced Institutes of Convergence Technology, Seoul National University, Suwon 443-270, Republic of Korea}
\begin{abstract}
Although redistribution of red blood cells at bifurcated vessels is highly dependent on flow rate, it is still challenging to quantitatively express the dependency of flow rate in plasma skimming due to nonlinear cellular interactions. We suggest a plasma skimming model that can involve the effect of fractional blood flow at each bifurcation point. For validating the new model, it is compared with \textit{in vivo} data at single bifurcation points, as well as microvascular network systems. In the simulation results, the exponential decay of plasma skimming parameter, $M$, along fractional flow rate shows the best performance in both cases.  

\end{abstract}
\maketitle

Red blood cells (RBCs) in microvessels are concentrated on the vessel core. Subsequently, a cell-free layer (CFL) with a few micrometer thickness is observed on the vessel wall. The CFL leads asymmetric redistribution of hematocrit at each bifurcation, called plasma skimming effect. As a continuous process of plasma skimming in microvascular networks, the average hematocrit in capillary beds is lower than the systemic hematocrit as reported in many previous studies \citep{cohnstein1888untersuchungen, johnson1971influence, schmid1975rbc, klitzman1979microvascular, lipowsky1980vivo, kanzow1981flow, kanzow1981analysis, sarelius1981microvascular}. Interestingly, the plasma skimming is recently revisited to develop new microchannels for detecting specific DNAs, proteins and cells by efficiently separating plasma from whole blood \citep{fan2008integrated, shevkoplyas2005biomimetic, kersaudy2013micro}. Also, it has been highlighted to accurately predict drug carrier distribution in the microvasculature \citep{lee2013near, tan2013influence, muller2014margination, lee2014quantifying, lee2016generalized, tan2016characterization, d2016microfluidic}. For utilizing the plasma skimming to new applications \textit{in vitro} and \textit{in vivo}, it is crucial to quantitatively predict the redistribution of RBCs and plasma at bifurcations. 

From the early 70s, several experiments for quantifying the plasma skimming were performed \citep{johnson1971influence, schmid1980cell, lipowsky1981microvessel, klitzman1982capillary, mchedlishvili1982effect, fenton1985nonuniform}. As pioneers, Pries et al \citep{pries1989red} measured plasma skimming regarding fractional blood flow in two different cases of \textit{in vivo} mouse model. The experiments confirmed previous studies that flow fractionation at the capillary entrance is an important determinant of capillary hematocrit, not the absolute flow velocity itself \citep{johnson1971influence}. Then, the plasma skimming was expressed by Logit model considering fractional flow rate and vessel diameters \citep{pries2005microvascular}. This model matches well with previous experimental data at single bifurcations with specific curve fitting parameters. Recently, for improving extensibility of plasma skimming model to various conditions in microvascular networks, Gould and Linninger \citep{gould2015hematocrit} suggested a new model that can quantify the plasma skimming with a single parameter, $M$. Also, \citet{lee2016generalized} introduced a generalized version of plasma skimming model for cells and drug carriers.  

In this paper, we aim to mathematically model fractional blood flow in a simple and generalized manner in order to computationally study its significance in plasma skimming, and also to accurately predict plasma skimming in the microvasculature. For this task, a recently developed plasma skimming model \citep{gould2015hematocrit} is taken, and extended to take into account the effect of fractional blood flow. This new model is then validated with experimental data at single bifurcation level, and also at microvascular network level. 

\begin{figure*}
    \centering
    \includegraphics[scale=1.0]{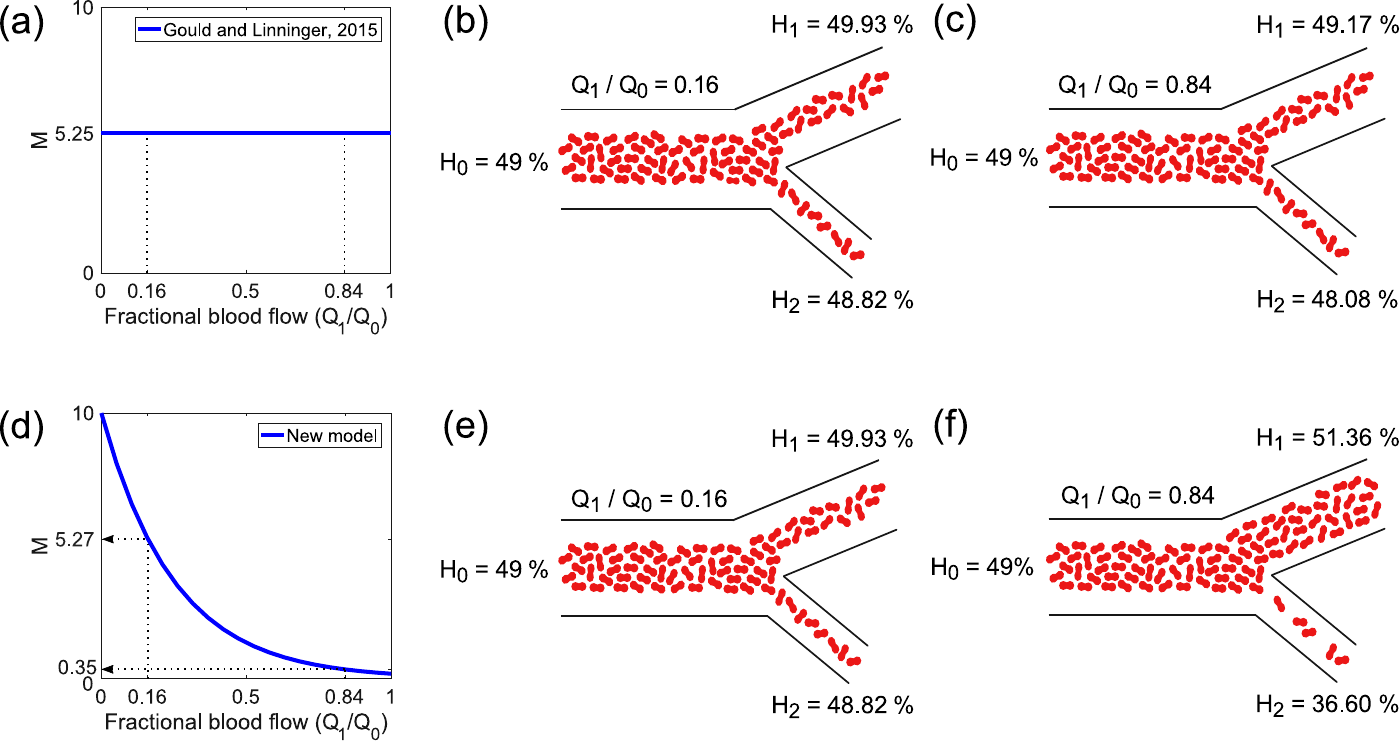}
	\caption{Plots of plasma skimming parameter $M$ against fractional blood flow, and
	illustrations of RBC redistribution in two cases. $Q_{1} / Q_{0}$ denotes
	the fractional blood flow between the largest daughter vessel and parent vessel. Without using the fractional blood flow model (a-c) there
	is negligible change in RBC redistributions at bifurcation since $M$ is set
	as a constant. When including the effect of fractional blood flow (d-f), both
	hemoconcentration and hemodilution after plasma skimming become more
	significant.} \label{fig1}
\end{figure*}

While there are other plasma skimming models \citep{pries2005microvascular,
guibert2010new}, the model developed by \citet{gould2015hematocrit} is
considered due to its easy extensibility. The model is as follows:
\begin{align}
H_1 &= H_0 - \Delta H = \zeta_1 H^* \\
H_2 &= \zeta_2 H^* \\
Q_0 H_0 &= Q_1 H_1 + Q_2 H_2 = Q_1 \zeta_1 H^* + Q_2 \zeta_2 H^* \\
\zeta_i &= \left( \frac{A_i}{A_0} \right) ^{\frac{1}{M}} \textrm{where}\ i = 1, 2
\end{align}

\noindent where $H$ is the hematocrit, $M$ is the plasma skimming parameter, $\zeta$ is
the hematocrit change coefficient due to plasma skimming, $Q$ is the flow rate,
$A$ is the cross-sectional area of each vessel, and subscript 0, 1, and 2
indicate the parent, and two daughter vessels, respectively. Specifically, the plasma skimming parameter $M$ is related to the cross-sectional distribution of RBCs near bifurcation. Small $M$ represents that RBCs are highly concentrated on the vessel core. In other words, plasma dominant region, or CFL, is developed on the near wall region. The two separated regions, expressed as RBCs and plasma areas, lead to strong plasma skimming. On the contrary, high $M$ means well-mixed RBCs and plasma. As a result, the plasma skimming effect will be diminished. Although plasma skimming is a function of hemodynamic parameters, $M$ was fixed at a constant value, $M$ = 5.25, for the entire microvasculature \citep{gould2015hematocrit} due to its complexity. 

Here, for improving the plasma skimming model, the flow rate change from parent to daughter vessels is expressed by 

\begin{equation}                                                                                           
M = M_{0} \cdot e^{-k \frac{Q_{1}}{Q_{0}}}
\label{eq:m}
\end{equation}

\noindent where $M_{0}$ and $k$ are constant values for quantifying $M$ as a function of fractional blood flow. In our simulation, $M_{0}$ and $k$ are
10 and 4, respectively. Note that the subscript $1$ denotes the daughter vessel
with the largest diameter. Conceptually speaking, $M$ can be considered as a ratio between RBC collision force and hemodynamic lift force at vessel wall. In this sense, if $Q_1/Q_0$ is low, hemodynamic lift force at the corresponding daughter vessel is low compared to RBC collision force, and hence high $M$ value is obtained. This leads to more uniform distribution of RBCs at bifurcation. On the contrary, high $Q_1/Q_0$ induces higher hemodynamic lift force, resulting in low $M$ value. Since, in this case, the RBCs are more likely to flow along vessel core region towards the daughter vessel with larger diameter, stronger plasma skimming effect is produced. Therefore, the exponential decay function of $M$ weakens the plasma skimming at low $Q_1/Q_0$ and vice versa. For instance, when $Q_1/Q_0$ is reduced, the natural flow tendency from parent vessel to daughter vessel with larger diameter is disturbed and then well-mixed at the bifurcation point. Under such circumstance, the hematocrit change from plasma skimming is small. On the other hand, at high $Q_1/Q_0$, the natural flow with CFL from parent vessel is prolonged to daughter vessel with larger diameter, leading to hematocrit redistribution.

Figure~\ref{fig1} depicts plots of $M$ against fractional blood flow and
schematic illustrations of RBC redistribution with computed hematocrit values
with and without fractional blood flow model at single bifurcation. For this
computation, hematocrit value at parent vessel is set to 49\% and diameters of
parent, and two daughter vessels are set to 20$\mu$m, 17.5$\mu$m, and
16.5$\mu$m, respectively. Figure~\ref{fig1}(a) shows the plot of $M$ over
fractional blood flow when $M$ is set as a constant \citep{gould2015hematocrit}. Figure~\ref{fig1}(b) and (c) show the RBC redistributions when $Q_{1} / Q_{0}$ is 0.16 and 0.84,
respectively. Since no relationship between the plasma skimming parameter and fractional
blood flow was established in the original model, $M$ remains as a constant. In
this case, change in RBC redistribution when varying $Q_{1} / Q_{0}$, for a
given parent vessel hematocrit, is negligible. 

\begin{figure*}
    \centering
    \includegraphics[scale=0.7]{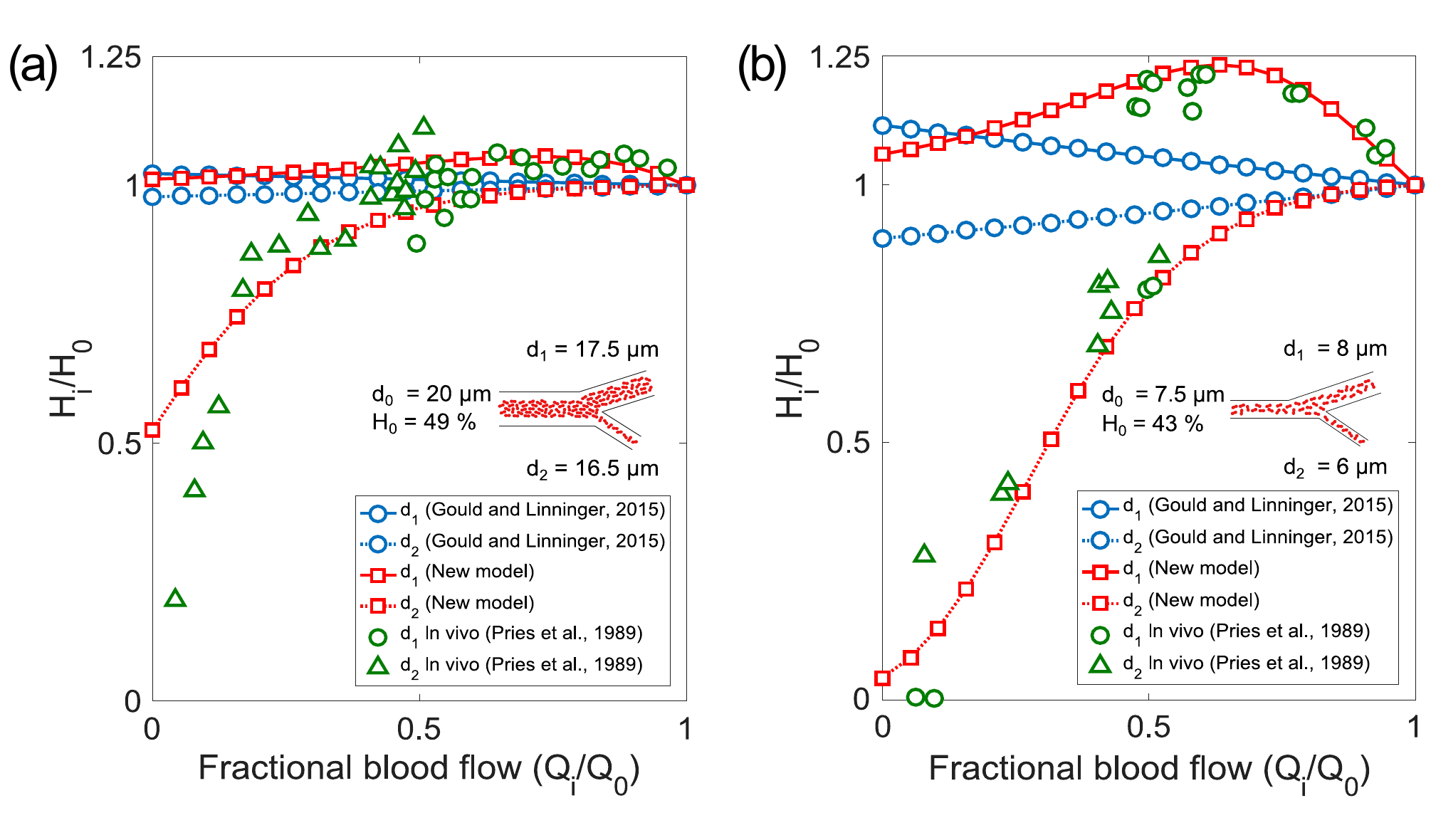}
	\caption{Ratio of hematocrit between parent and daughter vessels
	($H_{i} / H_{0}$) against fractional blood flow ($Q_{i} / Q_{0}$) at single
	bifurcation for comparing fractional blood flow model with the model
	developed by \citet{gould2015hematocrit} and experimental data \citep{pries1989red}. Two cases
	of geometries stated in Tab.\ref{tab:tab1} are considered. Significant
	amplifications in both hemoconcentration and hemodilution are produced by
	using fractional blood flow model, accurately matching with the experimental
	data.} \label{fig2}
\end{figure*}

Figure~\ref{fig1}(d) represents the plot of $M$ when Eq.\eqref{eq:m} is applied. As shown in Fig.\ref{fig1}(e), the hematocrit change at $Q_{1} / Q_{0}$ = 0.16 is similar to that in Fig.\ref{fig1}(b) due to the similar $M$ values. However, as described previously, increment in fractional blood flow produces greater plasma skimming, and hence $M$ is reduced. When $Q_{1} / Q_{0}$ is 0.84, $M$ now becomes 0.35. As depicted in Fig.\ref{fig1}(f), rather significant change is observed in RBC redistribution compared with Fig.\ref{fig1}(c), where the difference in hematocrit between the daughter vessels becomes more significant. By considering the effect of fractional blood flow, both hemoconcentration and hemodilution in
the daughter vessels are amplified. Equation~\eqref{eq:m} with corresponding
constants, as stated previously, is the only equation applied to model the
effect of fractional blood flow.

In order to validate the fractional blood flow model, plasma skimming at single
bifurcation is computed and compared with \textit{in vivo} experimental data
\citep{pries1989red}, along with the model developed by
\citet{gould2015hematocrit}. Logit model \citep{pries1989red, pries2005microvascular} is not
considered for single bifurcation since this model was developed based on curve
fitting of the same experimental data \citep{pries1989red}. The physiological
conditions as observed in the experiment are considered, and this is summarized
in Tab.\ref{tab:tab1}. $H_{0}$ denotes hematocrit value at parent vessel, and
$d_{0}$, $d_{1}$, and $d_{2}$ denote diameters of parent vessel, and two
daughter vessels, respectively. The same fractional model of Eq.\eqref{eq:m} is used
for both geometries.

\begin{table}[h]
	\begin{center}	
	\caption{Physiological conditions used for validation of fractional blood
	flow model at single bifurcation} \begin{tabular}{c c c c c}
		\hline
		Case&$H_{0}$&$d_{0}$&$d_{1}$&$d_{2}$\\ \hline
		1&49\%&20$\mu$m&17.5$\mu$m&16.5$\mu$m\\ 
		2&43\%&7.5$\mu$m&8$\mu$m&6$\mu$m\\ \hline
		\end{tabular}
	\label{tab:tab1}
	\end{center}
\end{table}

Figure~\ref{fig2} depicts the ratio of hematocrit, $H_{i} / H_{0}$, for two geometry cases from
Tab.\ref{tab:tab1}. The fractional blood flow model matches very well with the
experimental data, particularly in Fig.\ref{fig2}(b). While the model developed
by \citet{gould2015hematocrit} does not sufficiently capture hemoconcentration
and hemodilution, fractional blood flow model significantly amplifies them.
$H_{1}$ from fractional blood flow model on Fig.\ref{fig2}(b), for instance,
increases up to 1.25 showing significant hemoconcentration. Similarly, $H_{2}$
from fractional blood flow model shows very significant
hemodilution down to 0.04 as $Q_{2} / Q_{0}$ decreases. 

\begin{figure*}
    \centering
    \includegraphics[scale=0.9]{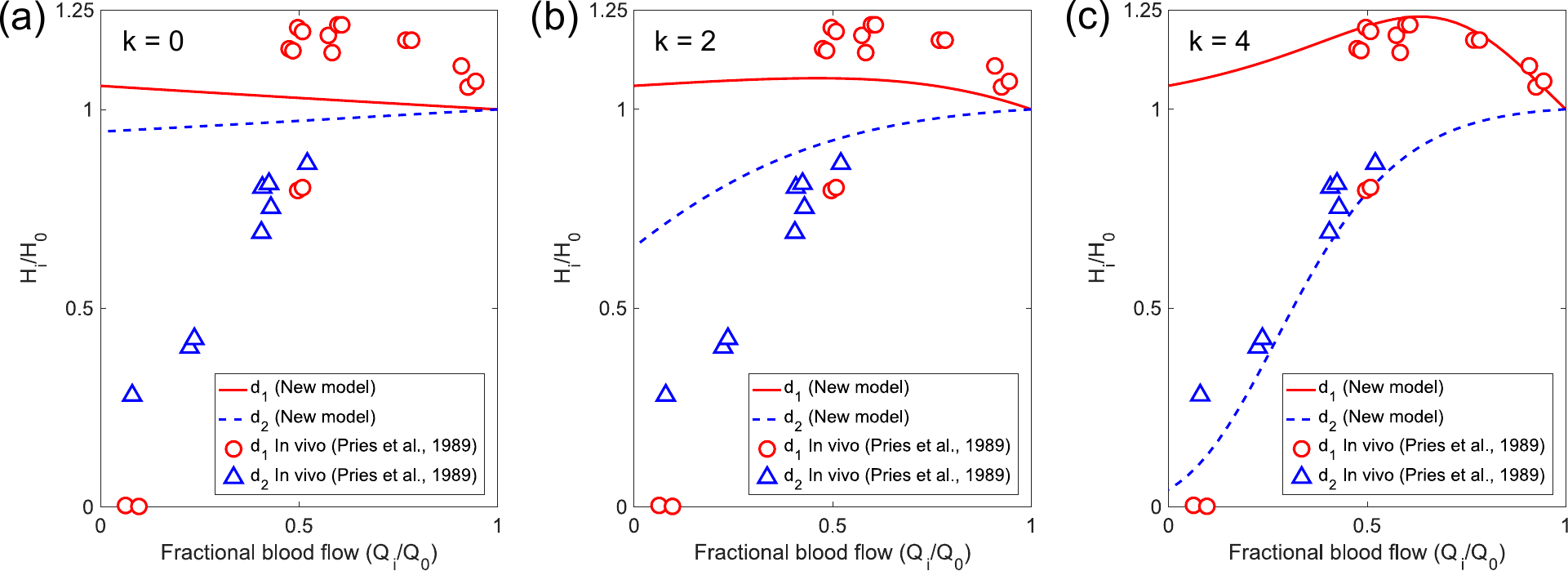}
	\caption{Ratio of hematocrit between parent and daughter vessels
	($H_{i} / H_{0}$) against fractional blood flow ($Q_{i} / Q_{0}$) at single
	bifurcation for different $k$ values. The second case stated in
	Tab.\ref{tab:tab1} are considered. The plots clearly show high sensitivity
	of $k$, and that $k=4$ gives the best match with the experimental data.}
	\label{fig3} \end{figure*}

Figure~\ref{fig3} depicts $H_{i} / H_{0}$ for different $k$ values. It must
be noted that $k$ is a highly sensitive parameter and must be chosen carefully,
and as Fig.\ref{fig3} shows, $k=4$ gives the best match with the experimental data.

\begin{figure*}[ht]
    \centering
    \includegraphics[scale=0.9]{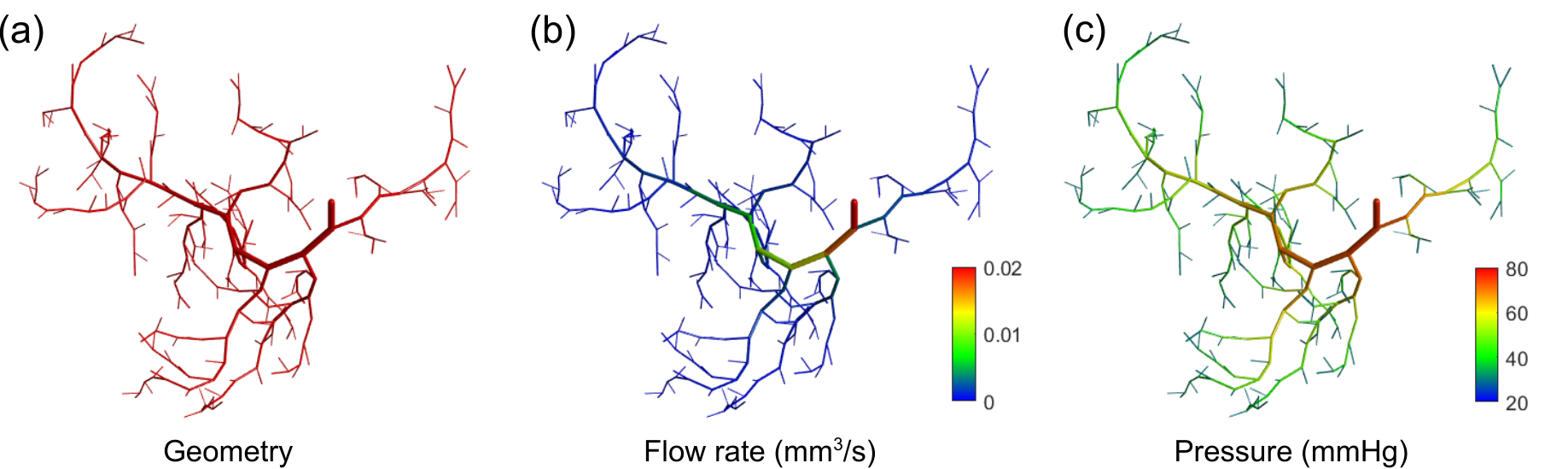}
	\caption{Computational model of microvascular network and corresponding hemodynamic calculations.} \label{fig4} 
\end{figure*}

\begin{figure*}
    \centering
    \includegraphics[scale=0.9]{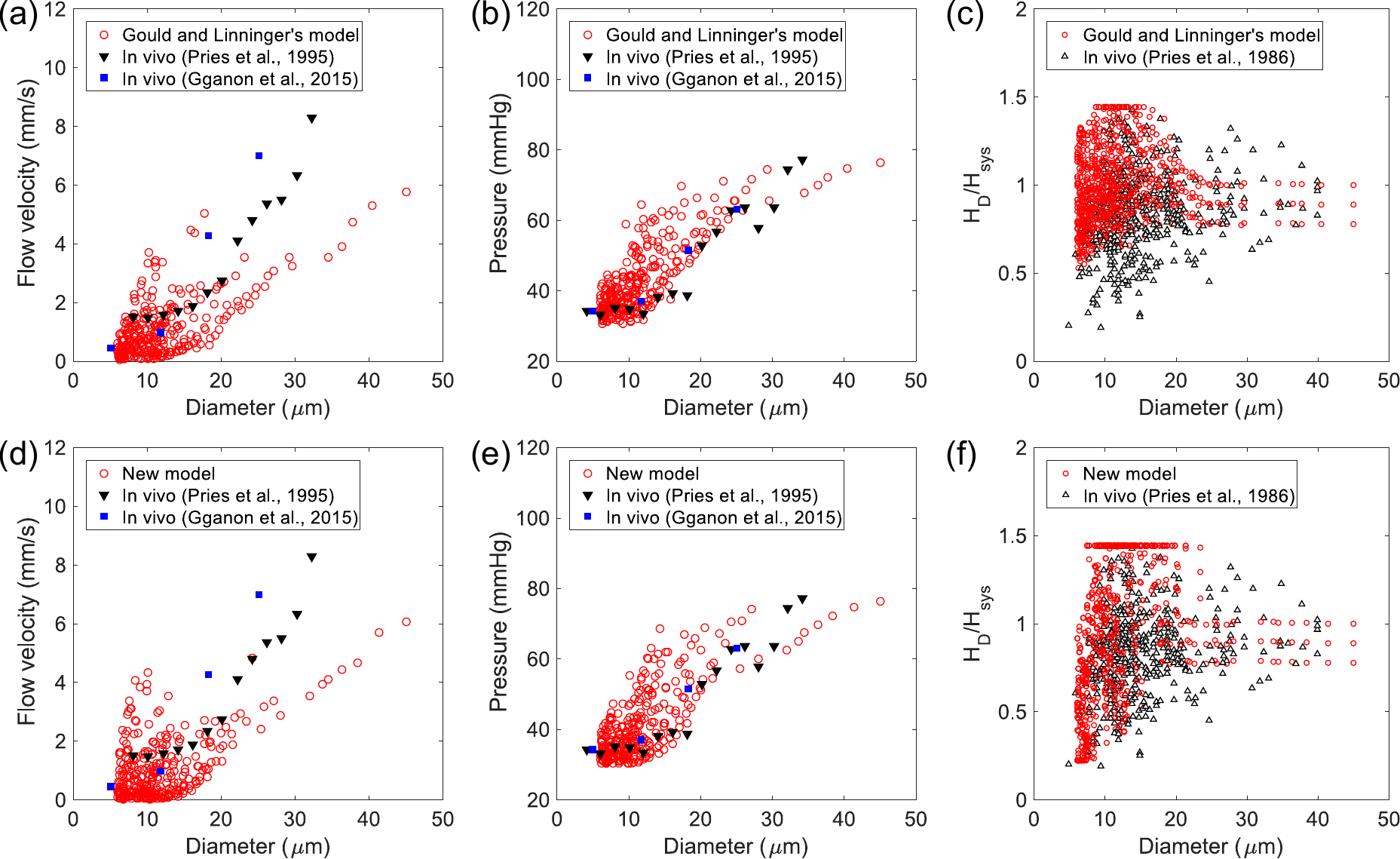}
	\caption{ Comparison of hemodynamic characteristics at microvascular network level between (a-c) Gould and Linninger's model and (d-f) Fractional flow model. Vessel diameters are asymmetrically decreased from 40 to 6$\mu$m. The pressure drop between the root vessel to capillary ends is set to 47mmHg. Flow velocity and pressure data are compared with two \textit{in vivo} experimental data \citep{pries1995structure, gagnon2015multimodal}. Also, relative hematocrit distribution along with vessel diameters at the microvascular network model. The systemic hematocrit, $H_{sys}$, is set to 0.45. Black triangle marks represent \textit{in vivo} experimental data \citep{pries1986generalization}. The initial hematocrit at the root vessel is varied from 0.3 to 0.45.} \label{fig5} 
\end{figure*}

To predict the plasma skimming effect at microvascular network level, the fractional blood flow model is coupled with a mathematical model of blood flow. A microvascular network model is computationally generated based on mathematical algorithms by choosing vessel diameter ($d_i$), vessel length ($l_i$), and bifurcation angles ($\theta_i$ and $\phi_i$)  \citep{lee2016generalized, yang2016predicting}. Diameters of daughter vessels at each bifurcation are governed by $d_0^{\gamma} = d_1^{\gamma} + d_2^{\gamma}$ where $\gamma$ is fixed at 3 \citep{murray1926physiological, sherman1981connecting}. Ratio of two daughter vessels, $\eta = d_2/d_1$, is utilized to control the geometric asymmetry of the entire microvascular network. With the diameter ratio, $\eta$, the diameters of daughter vessels are described by $d_{1} = \sqrt[\gamma]{d_{0}^{\gamma}/(1+N(\bar{\eta},\sigma^2)^{\gamma})}$ and $d_{2} = \sqrt[\gamma]{d_{0}^{\gamma}-d_{1}^{\gamma}}$ where $N$ is a normal distribution with mean $\bar{\eta} = 0.62$ and standard deviation $\sigma=0.1$ for capturing the heterogeneous diameter distribution. 

The diameter of root vessel is set to 40$\mu m$ and cut-off diameter is set to 6$\mu m$. Vessel lengths are governed by $l_i = \beta d_i^{n}$ where $\beta$ is 100 and $n$ is 0.46. 
Pressure drops between the root vessel and the capillary ends are 47mmHg. Flow rates of blood flow $(Q_i)$ is calculated by Poiseuille flow model, conservation of mass and $in~vivo$ viscosity laws \citep{pries1996biophysical} with the reference viscosity of plasma, fixed at $9\cdot 10^{-6}$ mmHg$\cdot$s \citep{yang2013design}. To express the variation of systemic hematocrit, the initial hematocrit values have a range from 0.3 to 0.45. The plasma skimming is controlled by considering CFL thickness in the plasma skimming model as $M'/M = 1+10e^{-100\delta'}$ where $\delta'$ is the relative CFL with respect to vessel diameter, which is determined by a curve-fitting of \textit{in vivo} experiment data \citep{tateishi1994flow},  $\delta' = (1.8e^{-6H}\sqrt{d-5.0}+0.5)/d$. The CFL function is applied in order to limit plasma skimming at highly RBC concentrated parent vessels. Since high hematocrit means very thin CFL and hence no plasma skimming, if CFL thickness is too small this function sets high $M$ to stop plasma skimming.

Figure~\ref{fig4} depicts the computationally generated microvasculature model and corresponding hemodynamic calculations. Systemic hematocrit of 45\% is applied as an initial boundary condition. Figure~\ref{fig4}(a) shows the microvasculature geometry used for predicting plasma skimming at microvascular network level. Figure~\ref{fig4}(b) and (c) are the computed flow velocity and pressure, respectively. 
Figure~\ref{fig5} shows the computed blood flow and hematocrit distribution along with vessel diameters. Two models, Gould and Linninger's model and the fractional blood flow model, are compared with \textit{in vivo} flow velocity and pressure data \citep{pries1995structure, gagnon2015multimodal}, and \textit{in vivo} hematocrit scatter data \citep{pries1986generalization}. As plotted in the figures, the mathematical model considering the effect of fractional blood flow holds good agreement with \textit{in vivo} data. From Fig.\ref{fig5}, one must note that application of flow rate ratio effect does not significantly alter both flow velocity and pressure. This is because the most significant parameter for determining flow velocity is vessel diameter as states Poiseuille's law. On the other hand, flow rate ratio do strongly influence hematocrit distribution since the most significant parameter for determining hematocrit is $\zeta$, which is governed by $M$. Therefore, no significant change in flow velocity and pressure is visible despite significant change in hematocrit distribution. Furthermore, numerous parameters at the network level correlate with microvascular transport of blood. Hence, the sensitivity analysis of plasma skimming model with flow rate dependency must be very carefully investigated at microvascular network level. In this paper, we aim to solely study the effect of fractional flow rate on plasma skimming. For this reason, same hematocrit cut-off conditions are applied to both models to compare the results under same circumstances: artificial hematocrit cut-off value of 1.5 and CFL function. Both models capture the plasma skimming in capillary beds. However, unlike Gould and Linninger's model which shows dense hematocrit distribution in capillary beds particularly below 15$\mu$m, fractional blood flow model gives more sparsely dispersed hematocrit even below 10$\mu$m. Such distribution is obtained due to amplification in plasma skimming effect induced by fractional flow rate.

In conclusion, for the first time, the effect of fractional blood flow on plasma skimming of RBCs in the microvasculature is mathematically designed, and quantitatively predicted. As shown from the results, the fractional blood flow model accurately matches with \textit{in vivo} experimental data, at both single bifurcation and microvascular network level, indicating that fractional blood flow is an important parameter that must be taken into account for studying plasma skimming. Furthermore, these results quantitatively validate previous qualitative and experimental studies that fractional blood flow greatly affects plasma skimming. 

This research was supported by Basic Science Research Program through the National Research Foundation of Korea (NRF) funded by the Ministry of Education (NRF-2015R1D1A1A01060992), and by the Bio and Medical Technology Development Program of the National Research Foundation (NRF) funded by the Ministry of Science, ICT and Future Planning (2016M3A9B4919711).

\bibliography{mybibfile}

\end{document}